\documentclass[%
 reprint,
superscriptaddress,
 amsmath,amssymb,
 aps,
]{revtex4-2}

\usepackage{amssymb}
\usepackage{amsthm}
\usepackage{amsmath}
\usepackage{geometry}                
\usepackage{algorithm}
\usepackage{algorithmic}
\usepackage{comment}
\usepackage{bm}
\usepackage{placeins}
\usepackage{graphicx}
\usepackage{xcolor}
\usepackage{siunitx}
\usepackage{color}
\usepackage{cleveref}
\usepackage{subcaption}
\usepackage{mathrsfs}
\usepackage{notoccite}
\usepackage{braket}
\usepackage{caption}
\usepackage[normalem]{ulem}

\def\r{\bm r}

\usepackage{mathtools}

\def\Ehal{E_{\rm HAL}}

\def\0{{\bm 0 }}

\def\posterior{\pi}

\newcommand{\mse}[1]{}
\newcommand{\msenew}[1]{}
\newcommand{\remove}[1]{}

\begin{document}


\title{Hyperactive Learning for Data-Driven Interatomic Potentials}

\author{Cas van der Oord}
\email{casv2@cam.ac.uk}
\affiliation{University of Cambridge, Cambridge, CB2 1PZ, U.K.}
\author{Matthias Sachs}
\affiliation{University of Birmingham, Birmingham, B15 2TT, U.K.}
\author{D\'{a}vid P\'{e}ter Kov\'{a}cs}
\affiliation{University of Cambridge, Cambridge, CB2 1PZ, U.K.}
\author{Christoph Ortner}
\affiliation{University of British Columbia, Vancouver, BC, V6T 1Z2, Canada}
\author{G\'{a}bor Cs\'{a}nyi}
\affiliation{University of Cambridge, Cambridge, CB2 1PZ, U.K.}

\date{\today}

\begin{abstract}

Data-driven interatomic potentials have emerged as a powerful class of surrogate models for {\it ab initio} potential energy surfaces that are able to reliably predict macroscopic properties with experimental accuracy. In generating accurate and transferable potentials the most time-consuming and arguably most important task is generating the training set, which still requires significant expert user input. To accelerate this process, this work presents \text{\it hyperactive learning} (HAL), a framework for formulating an accelerated sampling algorithm specifically for the task of training database generation. The key idea is to start from a physically motivated sampler (e.g., molecular dynamics) and add a biasing term that drives the system towards high uncertainty and thus to unseen training configurations. Building on this framework, general protocols for building training databases for alloys and polymers leveraging the HAL framework will be presented. 
For alloys, ACE potentials for AlSi10 are created by fitting to a minimal HAL-generated database containing 88 configurations (32 atoms each) with fast evaluation times of $<$100 $\mu$s/atom/cpu-core. These potentials are demonstrated to predict the melting temperature with excellent accuracy. For polymers, a HAL database is built using ACE, able to determine the density of a long polyethylene glycol (PEG) polymer formed of 200 monomer units with experimental accuracy by only fitting to small isolated PEG polymers with sizes ranging from 2 to 32.

\end{abstract}

\maketitle





\section{Introduction}

Over the last decade there has been rapid progress in the development of data-driven interatomic potentials, see the review papers \cite{deringer2019,deringer2021,keith2021,THOMPSON2015316, WANG2018178, Novikov2020-ly}. Many systems are often too complex to be modelled by an empirical description yet inaccessible to electronic structure methods due to prohibitive computational cost. Richly parametrised data-driven interatomic potentials bridge this gap and are able to successfully describe the underlying chemistry and physics by approximating the potential energy surface (PES) with quantum mechanical accuracy \cite{sosso2018, deringer_bernstein2021, kapil2022}. This approximation is done by regressing a $\textit{high-dimensional}$ model to training data collected from electronic structure calculations.

Over the years many approaches have been explored using a range of different model architectures. These include Artificial Neural Networks (ANN) based on atom centered symmetry functions~\cite{behler2007} and have been used in models such as ANI~\cite{ANI-2x, Smith2017ANI-1:Cost} and DeepMD \cite{Wang2018DeePMD-kit:Dynamics}. Another widely used approach is Gaussian Process Regression (GPR) implemented in models such as SOAP/GAP \cite{bartok2015g, Bartok2017MachineMolecules}, FCHL \cite{Christensen2020FCHLLearning} and sGDML \cite{Chmiela2019SGDML:Learning}. Linear approximations of the PES have also been introduced initially by using permutation invariant polynomials (PIPs) \cite{Braams2009PermutationallyDimensionality} and the more recent atomic PIPs variant~\cite{van2020regularised, Allen2021AtomicFields}. Other linear models include spectral neighbour analysis potentials \cite{thompson2015spectral} based on the bispectrum \cite{Bartok2010}, moment tensor potentials \cite{shapeev2016} and the atomic cluster expansion (ACE) \cite{drautz2019atomic, DUSSON2022110946, kovacs2021linear}. More recently, message passing neural network (MPNN) architectures have been introduced \cite{Schutt2017Quantum-chemicalNetworks, Anderson2019Cormorant:Networks, Unke2019PhysNet:Charges, Klicpera2020DirectionalGraphs,  schutt2021equivariant, gasteiger2021gemnet, batzner2022, batatia2022mace}
the most recent of which have been able to outperform any of the previously mentioned models regarding accuracy on benchmarks such as MD17 \cite{Chmiela2017MachineFields} and ISO17 \cite{Schutt2017SchNet:Interactions}. Central to all of these models is that they are fitted to a training database comprised of configurations $R$ labelled with observations comprising total energy $\mathcal{E}_{R}$, forces $\mathcal{F}_{R}$ and perhaps virial stresses $\mathcal{V}_{R}$, obtained from electronic structure simulations. By performing a regression on the training data model predictions $E$ of the total energy, and estimates of the respective forces $F_{i} = -\nabla_{i} E$ can be determined. Here, the $\nabla_{i}$ operator denotes the gradient with respect to the position of atom $i$.




Building suitable training databases remains a challenge and the most time consuming task in developing general data-driven interatomic potentials \cite{Rowe2020, GAP_Si, GAP_P}. Databases such as MD17 and ISO17 are typically created by performing Molecular Dynamics (MD) simulations on the structures of interest and selecting decorrelated configurations along the trajectory. This approach samples the potential energy surface according to its Boltzmann distribution. Once the training database contains sufficient number of configurations, a high dimensional model may be regressed in order to accurately interpolate its potential energy surface.  The interpolation accuracy can be improved by further sampling, albeit with diminishing returns.
However, it is by no means clear that the Boltzmann distribution is the optimal measure, or even a ``good'' measure, from which to draw samples for an ML training database. Indeed, it likely results in severe undersampling of configurations corresponding to defects and transition states, particularly for material systems with high barriers, which nevertheless have a profound effect on material properties and are often the subject of intense study.



A lack of training data in a sub-region can lead to deep unphysical energy minima in trained models, sometimes called ``holes'', which are well known to cause catastrophic problems for MD simulations: the trajectory can get trapped in these unphysical minima or even become unstable numerically for normal step sizes. A natural strategy to prevent such problems is active learning (AL): the simulation is augmented with a stopping criterion aimed at detecting when the model encounters a configuration for which the prediction is unreliable. Intuitively, one can think of such configurations as being ``far'' from the training set. When this situation occurs, a ground-truth evaluation is triggered, the training database extended, and the model refitted to the enlarged database. In the context of data-driven interatomic potentials, this approach was successfully employed by the linear moment tensor potentials~\cite{PODRYABINKIN2017171, GUBAEV2019148} and the Gaussian process (GP) based methods FLARE \cite{Vandermause2020OntheflyAL, vandermause2022} and GAP \cite{sivaraman2020} which both use site energy uncertainty arising from the GP to formulate a stopping criterion in order to detect unreliable predictions during simulations. 

The key contribution of this work is the introduction of the {\it hyperactive learning} \msenew{(HAL)} framework. Rather than relying on normal MD to sample the potential energy and wait until an unreliable prediction appears (which may take a very long time once the model is decent), we continually bias the MD simulation towards regions of high uncertainty. By balancing the physical MD driving force with such a bias we accelerate the discovery of unreliably predicted configurations but retain the overall focus on low energy regions that are important for modelling. This exploration-exploitation trade-off originates from Bayesian Optimisation (BO), a technique used to efficiently optimise a computationally expensive ``black box'' function \cite{brochu2010tutorial}. BO has been shown to yield state-of-the-art results for optimisation problems while simultaneously minimising incurred computational costs by requiring fewer evaluations \cite{wilson2018}. In atomistic systems BO has been applied in global structure search ~\cite{jorgensen2018, bisbo2022, merte2022,christiansen2022} where the PES is optimised to find stable structures. Other previous work balancing exploration and exploitation in data-driven interatomic potentials is also closely related, where configurations were generated by balancing high uncertainty and high-likelihood (or rather low-energy) \cite{koda2021}. Here the PES was explored by perturbing geometries while monitoring uncertainty rather than explicitly running MD. Note that upon the completion of this work, we discovered a closely related work that also uses uncertainty-biased MD\cite{UDD-Al}. The two studies were performed independently, and appeared on preprint servers near-simultaneously.

In BO an acquisition function balances exploration and exploitation, controlled by a biasing parameter. \msenew{In our hyperactive learning framework, the HAL potential energy surface $\Ehal$, \remove{Many acquisition functions exists, with one being the Lower Confidence Bound (LCB) which we adopt in this work to set up the HAL potential energy surface $\Ehal$,}}

\begin{equation}
    \Ehal := E - \tau \sigma
    \label{eq:ehal}
\end{equation}

\msenew{takes on a similar role. Here $E$ is the predicted potential energy and $\sigma$ is an uncertainty measure, which is provided by the modeller.} The {\it biasing strength}, represented by biasing parameter $\tau$, controls the exploration of unseen parts of the PES and needs to be carefully tuned in order for the HAL-MD trajectory to remain energetically sensible. An on-the-fly auto-tuning of $\tau$ is presented in the Methods section. The addition of a biasing potential, accelerating the exploration of relevant configurations, has a long history in the study of rare events and free energy computations, using adaptive biasing strategies such as meta-dynamics \cite{laio2002escaping,bussi2006equilibrium}, umbrella sampling \cite{marsili2006self,dickson2010free}, and similar methods (e.g., \cite{darve2001calculating,henin2004overcoming}). While the biasing force in these methods is implicitly specified by the choice of a collective variable, the direction of the biasing force in HAL is the result of the choice of the uncertainty measure $\sigma$.

We make the general HAL concept concrete in the context of the ACE ``machine learning potential'' framework~\cite{drautz2019atomic,DUSSON2022110946}, however, the methods we propose are immediate applicable to linear models and to Gaussian process type models, and are in principle also extendable to any other ML potential that comes with an uncertainty measure, including deep neural network models. In the context of linear ACE models, described in detail in the methods section, the site energy is defined as a linear combination of basis functions, 
\begin{equation}
    E_{i} = \mathbf{c} \cdot \mathbf{B}_{i}. \quad 
\end{equation}
and total energy, $E = \sum_i E_i = {\bf c} \cdot {\bf B}$ where ${\bf B} = \sum_i {\bf B}_i$.

The prediction of the uncertainty $\sigma$ can, for example, be obtained through the use of an ensemble. Different methods of setting up such ensembles for linear, GP or NN frameworks can be used, such as dropout \cite{srivastava2014}, or bootstrapping \cite{WEHRENS200035}.  In this work, we leverage the linearity of the ACE model and adopt a Bayesian view of the regression problem so that we are able to use  unbiased uncertainty estimation. The drawback analytical estimates of uncertainty is that often they are expensive to compute, which would preclude their evaluation at every MD time step, as needed by HAL. We circumvent this problem by setting up a committee based estimator for the unbiased Bayesian uncertainty measure, which yields an efficient algorithm with negligible overhead on top of ordinary MD.
Assuming an isotropic Gaussian prior on the model parameters and Gaussian independent and identically distributed (i.i.d) noise on observations, yields an explicit posterior distribution $\posterior({\bf c})$ of the parameters from which one can deduce the variance $\sigma_{E}^{2}$ of the posterior-predictive distribution of total energies,
\begin{equation} \label{eq:sigma}
\sigma_{E}^{2} = \frac{1}{\lambda} + \boldsymbol{B}^{T} \boldsymbol{\Sigma} \boldsymbol{B},
\end{equation}
where the covariance matrix $\boldsymbol{\Sigma}$ is defined as 
\begin{equation}
    \boldsymbol{\Sigma}^{-1} = \alpha \mathbf{I} + \lambda \mathbf{\Psi}^{T}  \mathbf{\Psi}. 
\end{equation}
Here, $\alpha, \lambda$ are hyperparameters whose treatment is detailed in the methods section, and $\mathbf{\Psi}$ is the corresponding design matrix of the linear regression problem and depends on the observations to which the ACE model is fitted.

The evaluation of the uncertainty or variance $\sigma_{E}^{2}$ in equation~\eqref{eq:sigma} is computationally expensive for a large basis $\mathbf{B}$; scaling as $O(N_\textrm{basis}^{2})$. To improve computational efficiency, $\sigma_{E}^{2}$ can be approximated by using an ensemble $\{ \mathbf{c}^{k} \}_{k=1}^{K}$ obtained by sampling from the posterior $\posterior(\mathbf{c})$ (see Methods for further details), resulting in 
\begin{equation} \label{eq:sigma:mc}
\begin{aligned}
\tilde{\sigma}_{E}^{2} &= \frac{1}{\lambda} +\frac{1}{K}\sum_{k=1}^{K} (E^{k} - \bar{E})^{2},
\end{aligned}
\end{equation}
where $\bar{E} = \bar{\bf c} \cdot {\bf B}$ with ${\bf \bar{c}}$ being the posterior mean of the posterior distribution whose closed form is provided in \eqref{eq:posterior:params} of the methods section.
This is computationally efficient to evaluate, requiring a single basis evaluation $\mathbf{B}$ followed by $K$ dot-products with the ensemble parameters.

Throughout the remainder of this article we will fix the choice of uncertainty measure in the definition of the HAL energy to be the standard deviation of the posterior-predictive distribution of energy as outlined above, i.e., $\sigma=\sigma_E$, which we approximate as $\tilde{\sigma}=\tilde{\sigma}_E$. From both a theoretical and modelling perspective, it would be of interest to consider other measures of uncertainty as biasing terms. 
Further discussion of this aspect is provided in the methods section.

Having introduced HAL-MD it remains to specify a stopping criterion that can be used to terminate the dynamics and extract new training configurations. 
To that end we introduce a {\it relative force uncertainty}, $f_i$, which is attractive from a modelling perspective, as for instance liquid and phonon properties require vastly different absolute force accuracy but similar relative force accuracy, typically on the order of 3-10$\%$. Given the model committee we introduced to define $\tilde{\sigma}$ we define 
\begin{equation}
    f_{i} = \frac{\frac{1}{K} \sum_{k=1}^{K}  \| F_{i}^{k} - \bar{F}_{i}  \|}{\|  \bar{F}_{i}  \| + \varepsilon },
    \label{eq:rel_force_err}
\end{equation}
where $\bar{F}_{i}$ is the mean force prediction. Further, $\varepsilon$ is a regularising constant to prevent divergence of the fraction, and to be specified by the user, often set to around 0.2 eV/\si{\angstrom}. During HAL simulations, $f_{i}$ provides a computationally efficient means to detect emerging local (force) uncertainties and trigger new ab initio calculations once it exceeds a predefined tolerance, 
\begin{equation}
    \underset{i}{\max} \, f_{i} > f ^{\textrm{tol}}.
\end{equation}
The specification of $f_{\textrm{tol}}$ is both training data and model specific, and often requires careful tuning to achieve good performance. Too low $f ^{\textrm{tol}}$ keeps triggering unnecessary ab initio calculations, whereas too high leads to generation of unphysical high energy configurations. To avoid manual tuning and aid generality, we normalise $f_{i}$ onto $[0,1]$ through the application of the softmax function $s(f_i)$, resulting in the new stopping criterion 
\begin{equation}
 \underset{i}{\max} \, \frac{\exp f_{i}}{\sum_{i} \exp f_{i}} > s^\textrm{tol},
\end{equation}
where we use the default tolerance $s^{\textrm{tol}}= 0.5$.


The paper is structured as follows. Following an initial discussion of the performance of the relative force error measure $f_{i}$, its ability to predict true error is investigated and its performance benchmarked by assembling a reduced diamond structure silicon database. Next, the HAL framework is used to build training databases for an alloy (AlSi10) and polymer (polyethylene glycol or PEG) from scratch and the ability of the resulting ACE models are able to accurately predict the AlSi10 melting temperature and PEG density are shown. 

\section{Results and Discussion}

\subsection{Filtering an existing training set}
%
%
Before illustrating the HAL algorithm itself, we first demonstrate the ability of the relative force error estimate $f_{i}$ in Eq.~\eqref{eq:rel_force_err} to detect true relative force errors. To that end, we will use \mse{\remove{the}this} estimator to significantly reduce a large training set while maintaining accurate model properties relative to the DFT reference. The database we use for this demonstration was originally developed for a Si GAP model \cite{GAP_Si} \mse{\remove{covering} and covers} a wide range of structures ranging from bulk crystals in various phases, amorphous, liquid and vacancy configurations. The filtering process builds a reduced database by starting from a single configuration and selecting configurations containing the maximum $f_{i}$ from the remaining test configurations. Iterating this process accelerates the learning rate and rapidly converges model properties with respect to the DFT reference. The models we train \mse{\remove{in this was}} are linear ACE models \mse{\remove{containing} that consist of}  basis functions up to correlation order $\nu$=3, polynomial degree 20, outer cutoff set to 5.5 \si{\angstrom} and inner cutoff set to the closest interatomic distance in the training database. An auxiliary pair potential basis was used using polynomial degree 3\mse{\remove{ and},} outer cutoff 7.0 \si{\angstrom} and no inner cutoff. The weights for the energy $w_{E}$, forces $w_{F}$ and virials $w_{V}$, which are described in detail in the Methods section, were set to 5.0/1.0/1.0. The size of the committees used to determine $f_{i}$ was $K=32$.


\subsubsection{Si diamond: error correlation and convergence}
\begin{figure}
    \includegraphics[width=1.0\columnwidth]{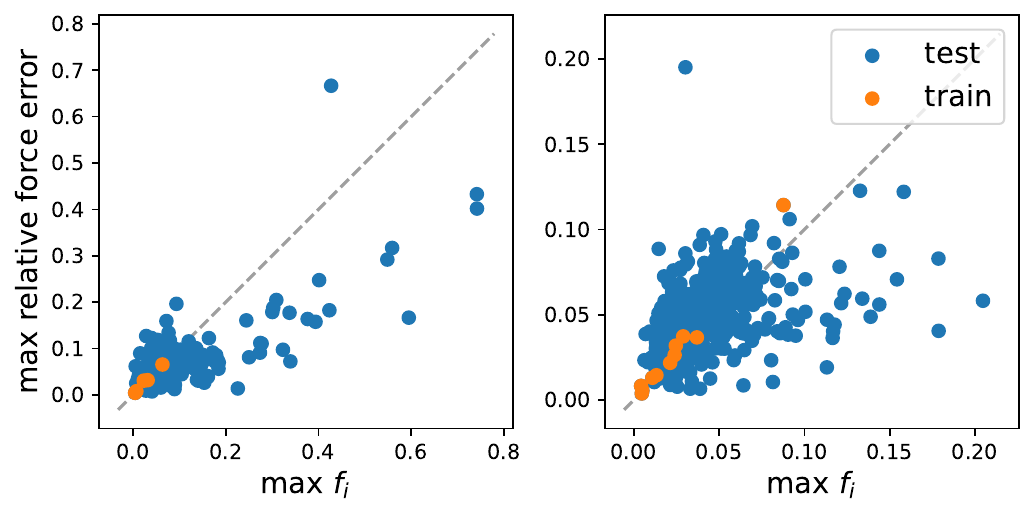}
    \includegraphics[width=1.0\columnwidth]{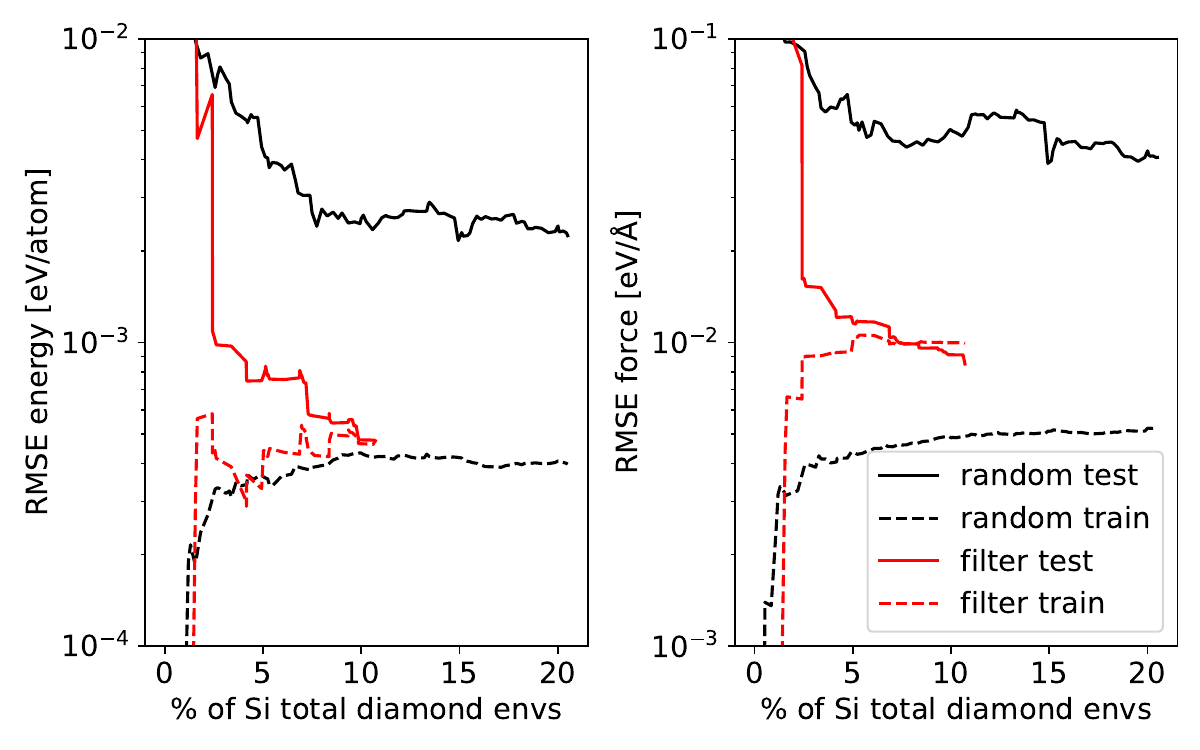}
    \caption{a) Maximum relative force error estimate $\max f_{i}$ versus error correlation plots for silicon diamond containing 4 and 10 training configurations. b) Learning rate comparison between filtering and random selection for silicon diamond.}
    \label{fig:HAL_filter_cor_conv_plot}
\end{figure}
Prior to training database reduction the ability of the relative force error estimate $f_{i}$ to predict relative force error is investigated. Fig.~\ref{fig:HAL_filter_cor_conv_plot}a compares the maximum relative force error in a configuration against the maximum of $f_{i}$ for two different training databases, containing 4 and 10 silicon diamond configurations respectively. The test configurations are the remaining configurations contained in the 489 silicon diamond configurations as part of the entire silicon database (totalling 16708 local environments). The regularising constant $\varepsilon$ was set to the mean force magnitude as predicted by the mean parameterisation. Both figures show good correlation between maximum relative force error and $\max \, f_{i}$, therefore making it a suitable criterion to be monitored during (H)AL strategies.

By leveraging the correlation of $f_{i}$ with true relative force error the existing silicon diamond database can be reduced by iteratively selecting configurations containing the largest relative force uncertainty as part of a greedy algorithms strategy. To demonstrate this\mse{,} a \mse{randomly selected} single configuration from the 489 silicon diamond configurations \mse{\remove{contained in}of} the silicon database was \mse{\remove{chosen first and}} fitted. Next, $f_{i}$ was determined over the remaining configurations and the configuration containing the largest $\max \, f_{i}$ \mse{was } added to the training database. This process was repeated\mse{\remove{ and the}. The} train and test error of this filtering procedure for silicon diamond is shown in Fig.~\ref{fig:HAL_filter_cor_conv_plot}b. It is benchmarked against performing random selection whereby, starting from the same initial configuration, test configurations were chosen at random from the pool of remaining \mse{\remove{test configurations} configurations of the training database}. The result indicates that $f_{i}$ accurately detects configurations with large errors and manages to accelerate the learning rate significantly relative to random selection. Good generalisation between training and test errors is achieved by using around 5$\%$ of the total environment contained in the original silicon diamond database.


\subsubsection{Si diamond: property convergence}

\begin{figure}
    \includegraphics[width=1.0\columnwidth]{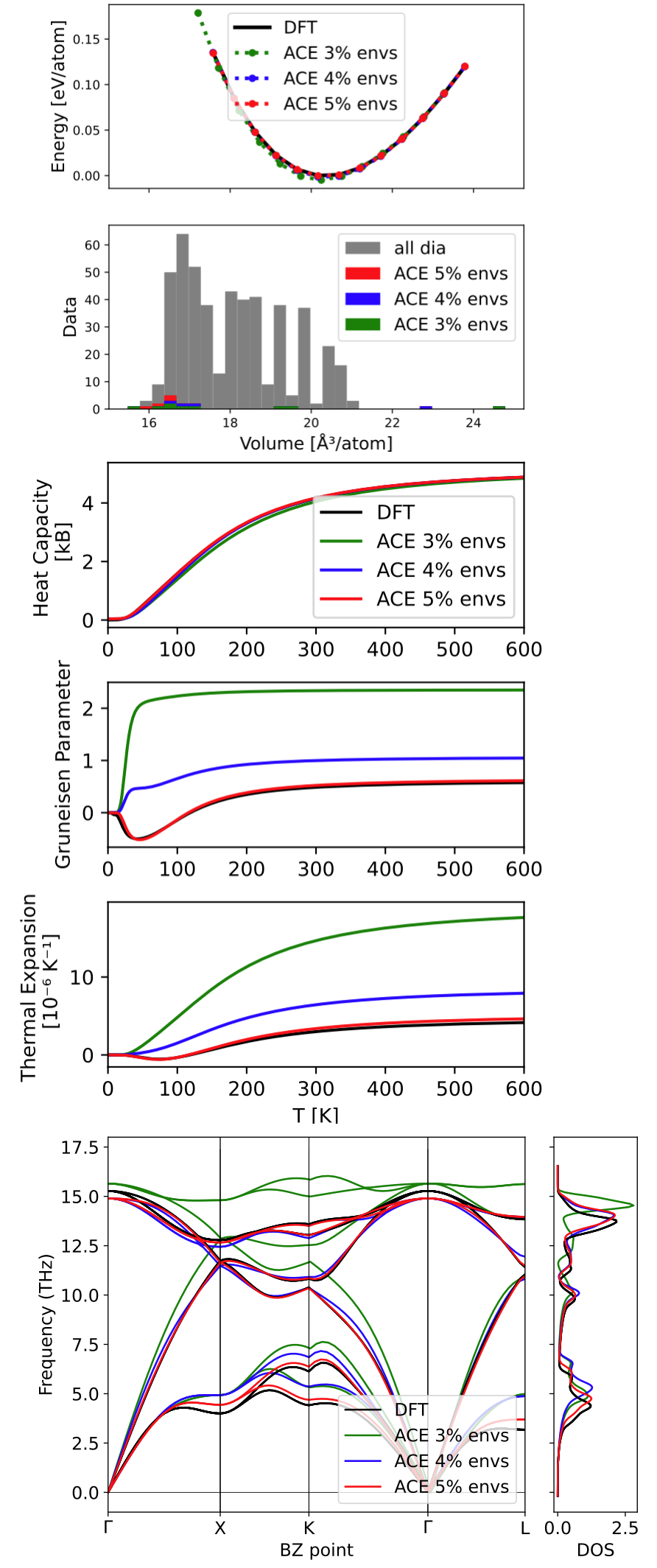}
    \caption{Property convergence for the energy volume (top), thermal properties (middle) and phonon spectrum (bottom) for filtering silicon diamond ACE models.}
    \label{fig:Si_prop_conv}
\end{figure}


The significant acceleration of the learning rate shown in Fig.~\ref{fig:HAL_filter_cor_conv_plot}b shows that generalisation between train and test error is rapidly achieved, in turn suggesting that property convergence is accelerated too. This is investigated \mse{\remove{by using the fitted linear ACE models during the filtering process and examining their predicted properties and comparing to the DFT reference.} by comparing macroscopic properties of the DFT reference with predictions of the ACE models that were fitted as part of the filtering process. }
These \mse{\remove{investigated}macroscopic} properties \mse{\remove{included:}include} elastic constants, energy volume curves, phonon spectrum and thermal properties for bulk silicon diamond. \mse{\remove{Three linear ACE models were chosen containing 3$\%$, 4$\%$ and 5$\%$ percent of the original total of silicon diamond environments, effectively slicing Fig.~\ref{fig:HAL_filter_cor_conv_plot}b along the x-axis. These models were fitted to 9 configurations (424 environments), 13 configurations (460 environments) and 17 configurations (608 environments) of silicon diamond respectively.
}
We report results of the ACE models that were fitted to 9 configurations (424 environments), 13 configurations (460 environments) and 17 configurations (608 environments), which, respectively, amount to approximately 3$\%$, 4$\%$ and 5$\%$ of silicon diamond environments contained in the original database.
}

Fig.~\ref{fig:Si_prop_conv} demonstrates that \mse{\remove{indeed}} property convergence for the energy volume curves, phonon spectrum and thermal properties are rapidly achieved by fitting to a fraction of the original database.
\mse{\remove{Fitting}In particular, fitting} to 5$\%$ of the original database reaches sufficient accuracy to describe all properties with good accuracy with respect to the DFT reference. This is again confirmed by elastic constants as predicted by the respective models as shown in Table.~\ref{tbl:si_table}. The convergence of the phonon spectrum in Fig.~\ref{fig:Si_prop_conv} is particularly \remove{is} noteworthy as relative errors on the order of a few percent on small forces $\sim$ 0.01 eV/\si{\angstrom} are \mse{\remove{required to be described accurately in order to describe the phonon spectrum well} typically required to accurately recover the phonon spectrum}. \mse{\remove{This is achieved be designing $f_{i}$ to consider relative force errors, which the greedy algorithm manages to detect and suppresses during the filtering process.} The fact that we achieve such small relative force errors while fitting on very few data points is a direct consequence of the design of the filter criterion $f_i$.}

\begin{table}
\centering
\begin{tabular}{l|l|l|l|l}
\hline
\hline
 &  $B$ &  $c_{11}$ & $c_{12}$ & $c_{44}$ \\
\hline
ACE 3\% envs & 98.2 & 188.1  & 53.3  & 79.7 \\
ACE 4\% envs & 84.2  & 159.8 & 46.4 & 75.7 \\
ACE 5\% envs & 82.5 & 148.7  & 49.3 & 73.7\\
\hline
DFT & 82.6 & 147.2 & 50.3 & 73.1 \\
\hline
\hline
\end{tabular}\\
\caption{Convergence of the elastic moduli (GPa) of the filtered ACE models relative to the CASTEP DFT reference.} 
\label{tbl:si_table}
\end{table}

\subsection{AlSi10}

\begin{figure*}
    \centering
    \includegraphics[width=1.0\textwidth]{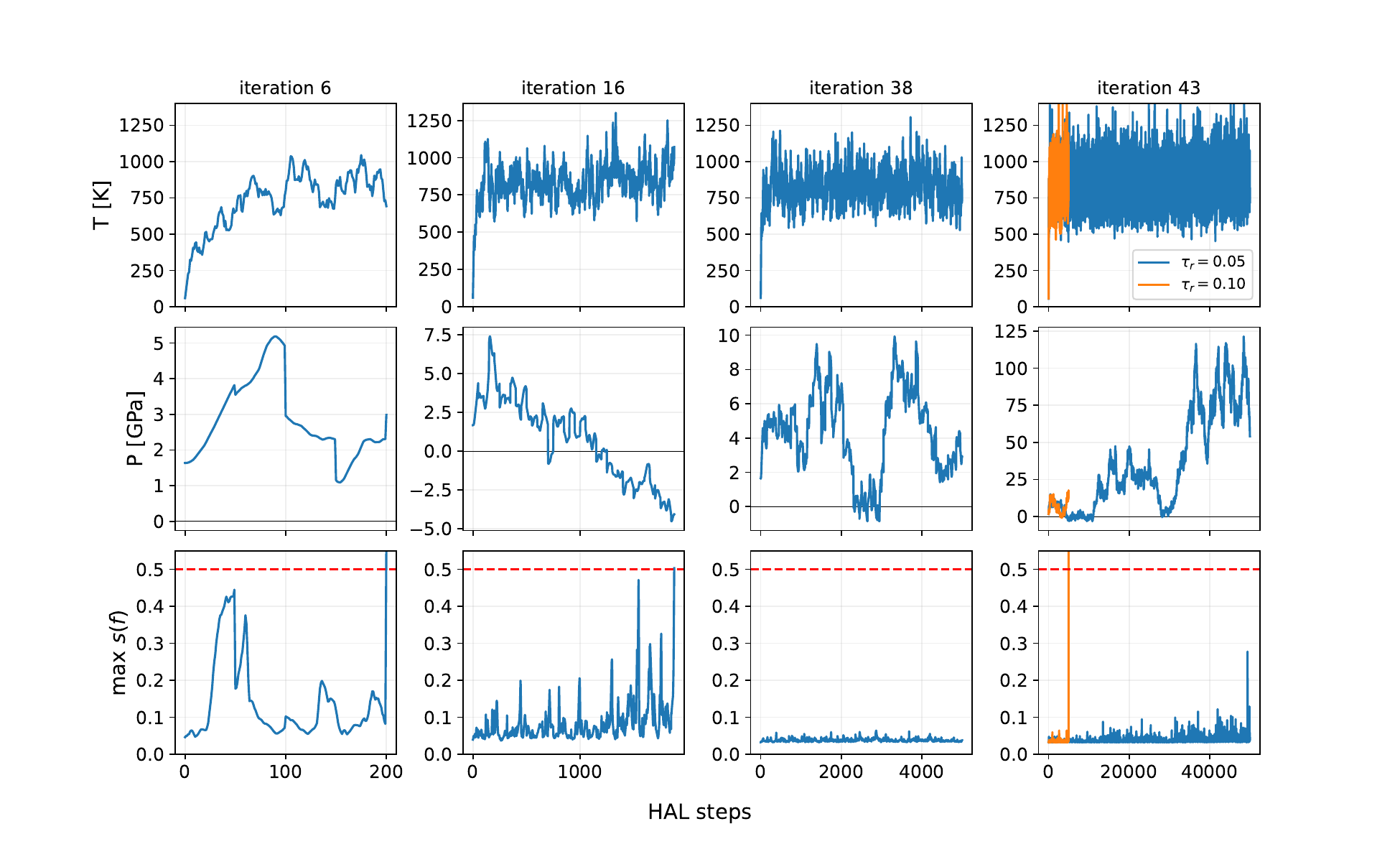}
    \caption{HAL dynamics for several iterations for the AlSi10 random alloy showing softmax normalised relative force error estimate $s(f)$, temperature and pressure. DFT calculations are triggered if the tolerance in red is reached. Pressure fluctuations are due to swap/volume MC steps on HAL potential energy surface $\Ehal$. }
    \label{fig:HAL_AlSi_solid}
\end{figure*}

\begin{figure*}
    \centering
    \includegraphics[width=0.8\textwidth]{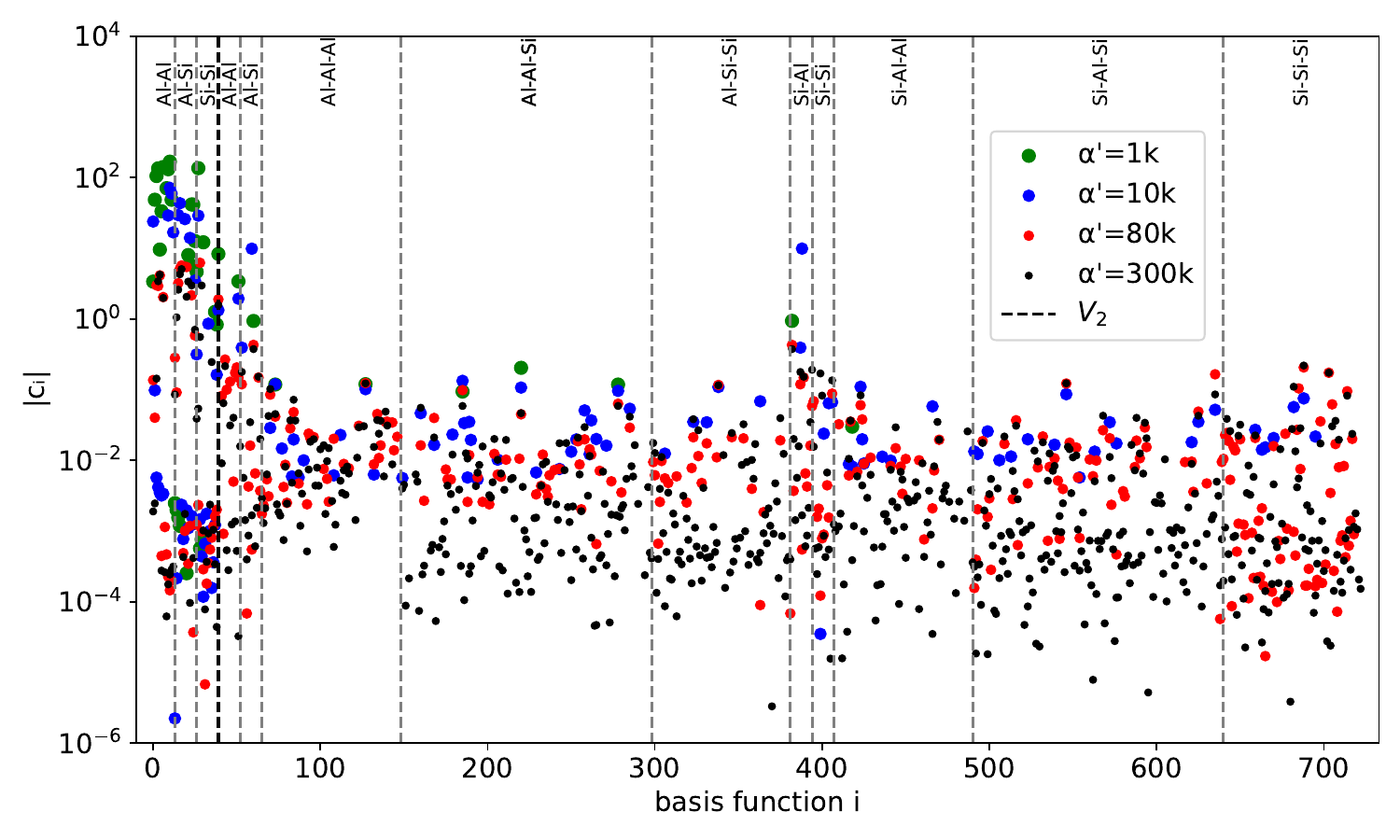}
    \caption{Coefficient magnitude $|\mathbf{c}_{i}|$ for the 723 basis functions grouped per correlation order and element interaction for various ARD tolerances $\alpha^{\prime}$. Large coefficients are assigned to pair interactions, partly captured by the auxiliary pair potential $V_{2}$, as most of the binding energy is contained in these interactions.}
    \label{fig:AlSi10_ARD}
\end{figure*}

\begin{figure*}
    \centering
    \includegraphics[width=0.8\textwidth]{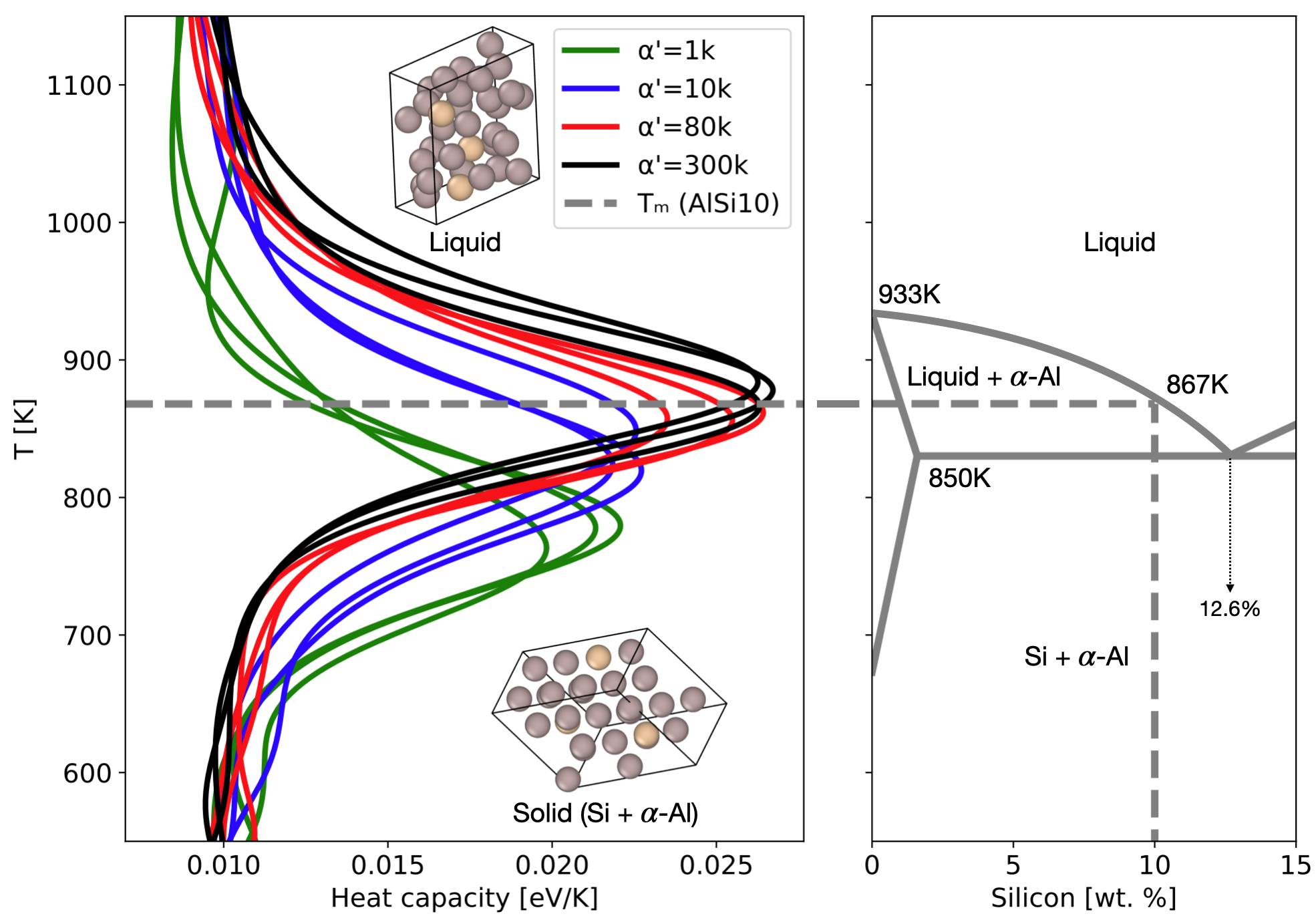}
    \caption{NS determined heat capacity for ARD fitted linear AlSi10 ACE models (left) and schematic phase diagram for AlSi10 \cite{AlSi_phase_diagram} (right). Excellent agreement with the melting temperature is demonstrated for fits with large $\alpha^{\prime}$. }
    \label{fig:AlSi_NS}
\end{figure*}

\begin{table*}
\centering
\begin{tabular}{c|c|c|c|cc|cc}
\hline
\hline
 $\alpha^{\prime}$ & $N_\textrm{basis}$ & Performance & Fit & \multicolumn{2}{c|}{Training Error} & \multicolumn{2}{c}{Test Error} \\
 &  & ($\mu$s/atom & (s) & E & F & E & F  \\
 &  & /core) & & (meV/at) & (eV/\si{\angstrom}) & (meV/at) & (eV/\si{\angstrom})  \\
\hline
  1$k$ &  38 &  62 &  2 &  7.693 & 0.135 & 8.006 & 0.147 \\
 10$k$ & 116 & 83 & 3 & 4.199 & 0.095 & 6.229 & 0.104 \\
 80$k$ & 295 & 85 & 17 & 2.401 & 0.080 & 5.131 & 0.089 \\
300$k$ & 621 & 99 & 63 & 1.869 & 0.074 & 5.188 & 0.095 \\
\hline
\hline
\end{tabular}\\
\caption{Train/test error splits for HAL generated AlSi10 database for varying ARD tolerance $\alpha^{\prime}$. Larger ARD tolerance $\alpha^{\prime}$ includes more basis functions but increases performance and fitting time. } 
\label{tbl:ARD_table}
\end{table*}


This section outlines the general HAL protocol for building training databases for alloys and demonstrates how an AlSi10 linear ACE model is built from scratch in an automated fashion. By using the relative force error estimate $f_{i}$ previously discussed as a stopping criterion to trigger ab initio evaluations it will be shown how an ACE model is created for AlSi10 using HAL. The HAL generated ACE model will be able to accurately model the liquid-solid phase transition and predict its melting temperature with excellent accuracy. The ACE models used in this section contained basis functions \mse{\remove{upto}up to correlation order} $\nu=2$ and polynomial degree 13 as well as an outer cutoff 5.5 \si{\angstrom}. The ACE inner cutoff was set to 1.5 \si{\angstrom} during the HAL stage of collecting data and moved towards the closest interatomic distance once all training data had been generated. An auxiliary pair potential $V_{2}$ added to aid stability also added to the basis including functions up to polynomial degree 13 and an outer cutoff of 6.0 \si{\angstrom}. The weights for the energy $w_{E}$, forces $w_{F}$ and virials $w_{V}$ were set to  \mse{\remove{15.0/1.0/1.0}15.0, 1.0, 1.0, respectively}. 

The HAL procedure of building ACE models for alloys starts by creating set of a random crystal structures manually, from which a random alloy and liquid alloy training database are built in an iterative fashion. Once sufficient data for both phases has been collected, the HAL solid and liquid databases are afterwards combined in order to create a model \mse{that} accurately \mse{\remove{described}describes} both phases. The first step in the HAL protocol is the creation of a set of small initial random alloy database, which was formed of 32-atom FCC lattice configurations populated with 29 Al and 3 Si atoms, equivalent to 9.7 weight percent Si. This initial random alloy starting database contained 10 configurations with lattice constants ranging from 3.80 \si{\angstrom} to 4.04 \si{\angstrom} and was evaluated using CASTEP \cite{CASTEP} DFT. The main parameters chosen are as follows: plane-wave cutoff 300 eV, kpoint spacing 0.04 \si{\angstrom}$^{-1}$, 0.1 eV smearing, Pulay density mixing scheme and finite basis correction. 

An adaptive biasing parameter $\tau_{\textrm{r}}$=0.05 was chosen (for explicit definition see Methods section) and the temperature set to $T_{\textrm{solid}}$=800K in order to build the random solid alloy database starting from the 10 initial structures previously described. Besides running biased dynamics, HAL performed cell volume adjusting (adding Gaussian noise to cell vectors) and element swapping Monte Carlo (MC) steps during the simulation on the HAL potential energy surface $\Ehal$. These steps were accepted or rejected according to the Metropolis-Hastings algorithm \cite{greenberg1995}.

During HAL dynamics the softmax normalised relative force estimate $s(f)$ is evaluated and a ground-truth evaluation triggered once a predefined tolerance of $s^{\textrm{tol}}$=0.5 is met. A total of 42 HAL configurations were sampled as the HAL dynamics at this stage was stable for 5000 steps\mse{.\remove{reliably.}} The pressure $P$, temperature $T$ and $f_{i}$ are shown in Fig.~\ref{fig:HAL_AlSi_solid} for four \mse{exemplary} iterations with the first three being included in the training database, e.g. below or equal to iteration 42. The strong oscillations in the pressure $P$ are due to the volume and element swapping MC steps being accepted. \mse{\remove{A fourth iteration, referred to as iteration 43, is shown on the far right demonstrating that increasing the biasing strength to $\tau_{\textrm{r}}$=0.10 accelerates the HAL dynamics to failure more rapidly accelerating the discovery of uncertain configurations exhibiting large relative force errors and resulting model failure.
} Finally, as demonstrated in the case of the 43th HAL iteration that increasing
the biasing strength to $\tau_{\textrm{r}}$=0.10 results in a drastic acceleration (by a factor of 10) in the discovery of configurations with large relative force error.
}

\mse{\remove{Increasing the temperature to $T_{\textrm{liquid}}$=3000K a liquid random alloy training database was assembled using the HAL generated random alloy training database as initial starting configurations.}
Next, HAL was employed to assemble a database of liquid random alloys. HAL trajectories were initialized at configurations sampled by cycling through the training database of random solid alloys obtained in the previous HAL run. 
} HAL trajectories were simulated using a Langevin thermostat targeting a temperature regime of $T_{\textrm{liquid}}$=3000K, and and a proportional control barostat targeting pressure level of 0.1 GPa. No volume or swap MC steps were performed.

After generating generating 46 liquid alloy configurations using HAL\mse{,} the \mse{HAL} dynamics \mse{\remove{was}were reliably} stable for 5000 steps \mse{\remove{reliably}} and \mse{\remove{HAL dynamics terminated.} the database assembly for this temperature regime was terminated.}

\mse{\remove{Combining the 42 HAL generated random alloy configurations and 46 HAL generated liquid configurations formed the complete training database used to create AlSi10 linear ACE models.}
Finally, the 42 HAL generated random alloy configurations and 46 HAL generated liquid configurations were combined to a training database. This training base was used to fit linear ACE models for AlSi10 using Automatic relevance determination (ARD); see section \ref{sec:methods:ard} for details. 
}
\mse{\remove{
Using the previously described ACE parameters the design matrix $\mathbf{\Psi}$ assembled shaped $N_{\textrm{obs}} \times N_{\textrm{basis}}$ was 9064 $\times$ 723. Fitting was performed using ARD regression with various different thresholds $\alpha^{\prime}$ and details are } We considered various thresholds $\alpha^{\prime}$ for the pruning of model parameters. The performance of the pruned models in terms of computational speed, training and test errors, are} shown in Table \ref{tbl:ARD_table}. \mse{\remove{Both the contents and shape of the design matrix $\mathbf{\Psi}$ as well as ARD are discussed in the Methods section.}} The test set \mse{used to compute test error consisted of 14 solid and 14 liquid configurations. These configurations were obtained by sampling from the corresponding temperature and pressure regimes of the surrogate models obtained at the end of the two preceding HAL runs.} was assembled by continuing HAL iterations for both the random alloy and liquid\mse{\remove{and contained 14 HAL solid and 14 HAL liquid configurations.}}. Increasing $\alpha^{\prime}$ lowers the relevance criterion for the linear ACE basis functions in turn decreasing sparsity. A clear trade-off between sparsity and training error can be seen in Table \ref{tbl:ARD_table} which also includes model evaluation performance and fitting times. Increasing $\alpha^{\prime}$ not only decreases training error but also test \mse{\remove{set}} error up to $\alpha^{\prime}=300k$ for which the test  \mse{\remove{set}} error increases, a sign of overfitting. Due to the relatively small training database size \mse{\remove{the fitting time} the computing time to fit the models} remains low, around a minute or less using 8 threads on Intel(R) Xeon(R) Gold 5218 CPU @ 2.30GHz. Performance testing was done using LAMMPs and the PACE evaluator \cite{lysogorskiy2021performant} on the exact same processor and show that all models are within 100 $\mu$s/atom/core per MD step.


Further analysis of the ARD fitted models was done by examining the absolute value of the coefficients $|\mathbf{c}_{i}|$. \mse{\remove{ARD performs feature selection by determining the relevance of the basis functions and weighting them accordingly in the regression.}} Basis functions \mse{\remove{with a relevance}whose estimated prior precision is} below the predefined threshold are pruned away as can be seen in Fig.~\ref{fig:AlSi10_ARD}. Large coefficients are given to the pair interactions described by the auxiliary basis $V_{2}$ and two-body components of the ACE basis for all models, which is intuitive as most binding energy is stored in these pair interactions. Increasing $\alpha^{\prime}$ results in more (less relevant) basis functions being included with relatively smaller coefficients. For $\alpha^{\prime}=300k$ many of these low relevance coefficients of around $10^{-4}$ are included in the fit indicating a degree of overfitting - as confirmed by the test set error increase in Table \ref{tbl:ARD_table}. 

Next\mse{,} the melting temperature for each of the previously ARD fitted AlSi10 ACE models is determined. This was done using Nested Sampling (NS) which approximates the partition function of an atomic system by exploring the potential energy surface over decreasing energy (or enthalpy) levels, in turn determining the cumulative density of states \cite{Partay2021}. From the partition function any thermodynamic quantity can be derived such as the heat capacity. The heat capacity exhibits a signature peak for first-order phase transitions, which includes the liquid-solid transition occurring at the melting temperature. Extensive previous work has shown that NS is an accurate and reliable method for determining the melting temperature \cite{baldock2016determining, partay2018performance}. As it explores the entirety of configurational space including gas, liquid and solid phases NS also serves as an excellent test for model robustness. This robustness is partly achieved by the addition of the auxiliary pair potential previously described as $V_{2}$ which is added in order to ensure close range repulsion between atoms. 

\mse{\remove{The NS simulations were carried using 896 walkers formed of 32 atom cells (29 Al and 3 Si) using the PYMATNEST software \cite{PYMATNEST}.} The NS simulations where performed using the PYMATNEST software \cite{PYMATNEST} and 896 walkers each corresponding to a configuration of 32 atom containing 29 Al and 3 Si atoms.} Starting from the gas phase (initial cell volume of 500 \si{\angstrom}$^{3}$/atom) the walkers explored the potential energy surface by iteratively cloning and decorrelating walkers at decreasing enthalpy levels, passing through the liquid phase and ending up at the ground structure. The decorrelation was done by running MD for 6 timesteps using a 0.1 fs timestep and a total of 1024 MC steps by changing the cell volume, shearing/stretching the cell and swapping atoms (ratio 6:6:6:6) were performed. The NS cell pressure was set to 0.1 GPa and the minimum aspect ratio of the cell set to 0.85. By summing the enthalpies from the NS simulation the constant pressure partition function is determined from which the heat capacity can be derived through postprocessing. Three independent runs for \mse{\remove{the}} each of the ARD models \mse{that were} fitted to the AlSi10 HAL database were performed. \mse{The such obtained heat capacity curves are} shown in Fig.~\ref{fig:AlSi_NS}. All models predicted the expected FCC ground structure, but a difference in \mse{the} predicted melting temperature for varying $\alpha^{\prime}$ can be seen. Only the $\alpha^{\prime}=300k$ and $\alpha^{\prime}=80k$ models accurately \mse{\remove{determine}predict} the melting temperature of 867K as predicted by Thermo-Calc with the TCAL4 database \cite{Tang2016}. \mse{\remove{Comparing to}Comparison with} Table~\ref{tbl:ARD_table}

\subsection{Polyethylene glycol (PEG)}

\begin{figure*}
    \centering
    \includegraphics[width=0.8\textwidth]{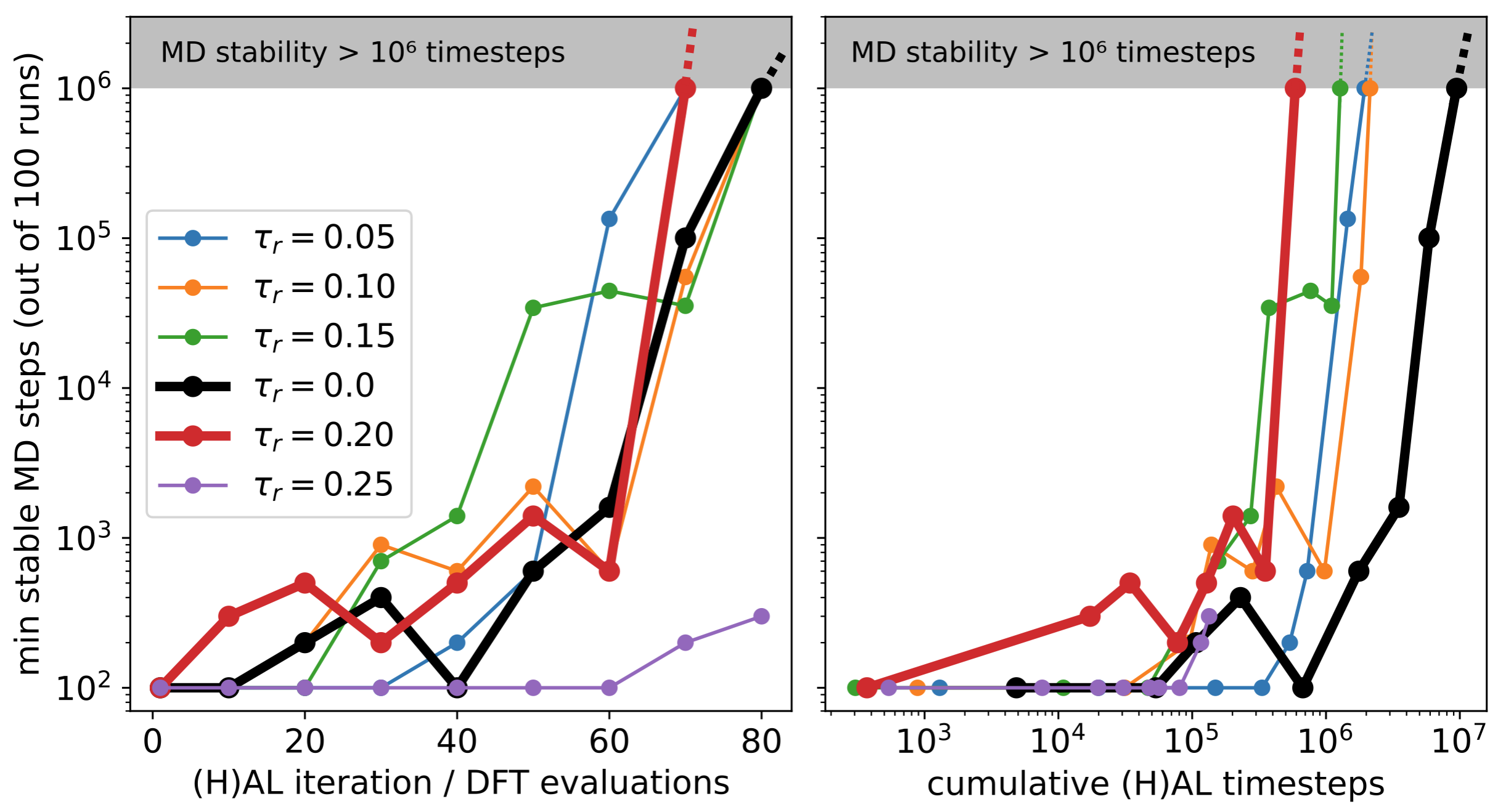}
    \caption{HAL vs AL benchmark comparing MD stability for 1 million MD steps over 100 seeds. Turning on biasing (non-zero $\tau_{\textrm{r}}$) creates ACE models achieving stable 100 million MD timestep faster than standard AL by up to an order of magnitude. }
    \label{fig:PEG_AL_HAL}
\end{figure*}

This section presents the application of HAL to build databases for polymers. Polyethylene glycol (PEG) has the formula H\big[OCH\textsubscript{2}CH\textsubscript{2}\big]\textsubscript{n}OH, where $n$ is the number of monomer units \cite{KARIMI20132465}. From a modelling perspective these polymers are challenging to simulate in vacuum as they form configurations ranging from tightly coiled up to fully stretched out structures. Due to the OH group at the end the polymer can also exhibit hydrogen bonding, which further complicates its description. These hydrogen bonds typically correspond to low energy configurations and are frequently formed and broken during long MD simulations. This section first presents a benchmark of HAL against AL followed by a demonstration HAL finding configurations exhibiting large errors. Finally, the potential fitted to small polymer units in vacuum is used to predict the density of a long PEG($n$=200) polymer in bulk with excellent accuracy relative to experiment. All DFT reference calculations in this section are carried out with the ORCA code~\cite{ORCA} using the $\omega$B97X DFT exchange correlation functional~\cite{ORCA_ex_cor_func} and 6-31G(d) basis set.

\subsubsection{PEG($n$=2): HAL vs AL}

In order to test whether HAL accelerates training database assembly relative to standard AL, a benchmark test was performed. An initial database containing 20 PEG($n$=2) polymers was created by running 500 K NVT molecular dynamics simulation using the general purpose ANI-2x forcefield \cite{ANI-2x} sampling every 7 ps. 
This database was fitted using an ACE basis containing basis functions up to correlation order $\nu$=3 and polynomial degree 10 with an outer cutoff 4.5 \si{\angstrom} and inner cutoff 0.5 \si{\angstrom}. The auxiliary pair potential basis up to polynomial degree 10 and outer cutoff 5.5 \si{\angstrom} and did not have an inner cutoff. The weights for the energy $w_{E}$, forces $w_{F}$ were set to 15.0 and 1.0 and remain constant throughout this seciton on PEG. AL (non-biasing, or $\tau$=0.0) and HAL simulations with varying biasing strengths $\tau_{\textrm{r}}$ were performed using a timestep of 0.5 fs at 500K. Configurations were evaluated using ORCA DFT once $s^{\textrm{tol}}$=0.5 was reached. 

The linear ACE models generated during the AL/HAL simulations were saved and subsequently used in a regular MD stability test and ran for 1 million MD steps at 500 K using a 1 fs timestep for 100 separate runs. A MD simulation was deemed stable if the CC and CO bonds along the chain where within 1.0-2.0\si{\angstrom} and the CH and OH bonds within 0.8-2.0\si{\angstrom} during the simulation. The {\it minimum} number of stable MD timesteps out of the 100 different simulations is shown in Fig.\ref{fig:PEG_AL_HAL} and demonstrates that up to $\tau_{\textrm{r}}$=0.20 a total of 80 (H)AL iterations are required in order to achieve a minimum MD stability of 1 million steps. The large biasing strength of $\tau_{\textrm{r}}$=0.25 results in unstable MD dynamics as too strong biasing causes the generation of exceedingly high energy configuration far away from the desired potential energy surface to be included in the training database. Fitting to these configurations leads to a poorly performing model as many unphysical configurations enter the training database resulting. 

The HAL run using a biasing strength of $\tau_{\textrm{r}}$=0.20, achieves minimum 1 million step MD stability after an order of magnitude fewer exploratory MD timesteps compared to standard AL. 

\subsubsection{PEG($n$=4): rare events}

Using PEG($n$=4) polymers this section will investigate the ability of HAL to generate and detect configurations with large errors. First a training database was built using the general purpose ANI-2x forcefield \cite{ANI-2x} at 500K and 800K using a timestep of 1 fs. Configurations were sampled every 7000 timesteps (7 ps), and used to assemble 500K and 800K databases. The 500K database was divided into 750 train configurations and 250 test configurations. The 800K training and test databases both contained 250 configurations. 
The linear ACE model was extended to include basis functions up to 12 for both the ACE and pair potential, while keeping the cutoffs and correlation order the same ($\nu$=3) too compared to the previous section on PEG($n$=2). 

\begin{figure*}
    \centering
    \includegraphics[width=1.0\textwidth]{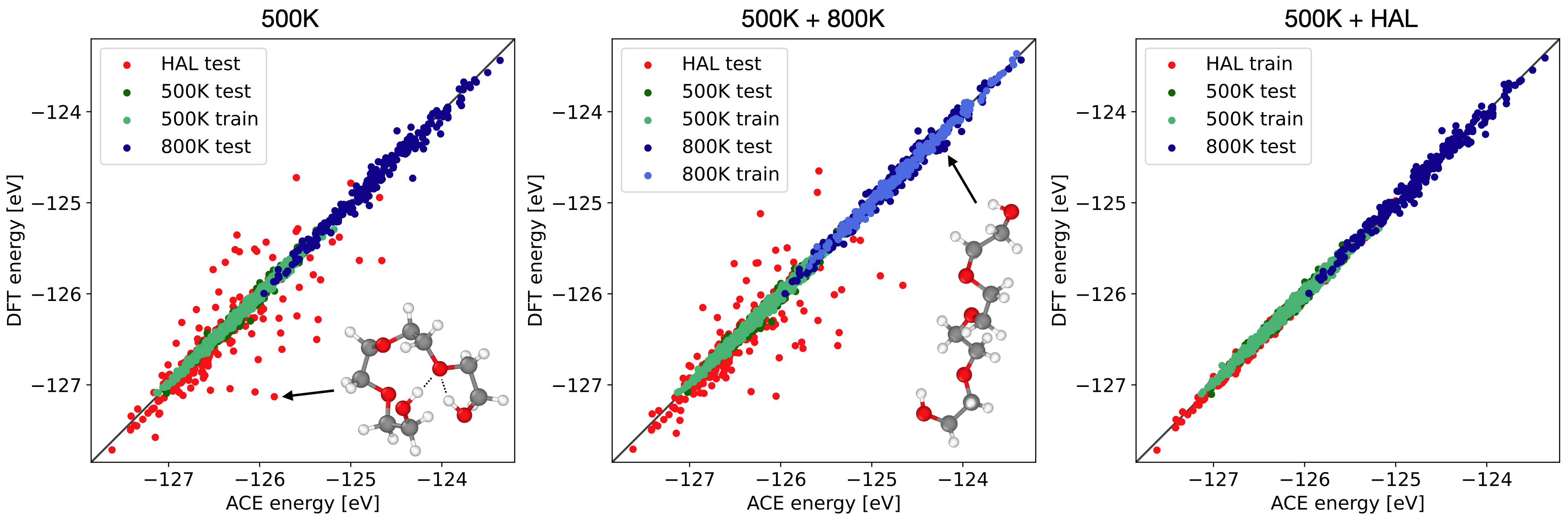}
    \caption{Energy scatter plots for the 500K (left), 500K+800K (middle) and 500K+HAL (right) ACE models. HAL configuration mostly exhibit (double) hydrogen bonding, or rare events, not contained in the MD 500K/800K decorrelated samples.}
    \label{fig:PEG_E_scatter}
\end{figure*}

Using the 500K MD sampled training database HAL was started using $\tau_{\textrm{r}}$=0.10 and a timestep of 0.5 fs. The stopping criterion $s^{\textrm{tol}}$ set to 0.5. A total of 200 HAL configurations were generated and formed a HAL database used for both a train and test set. Using the previously described basis three models were created fitted to: 500K, 500K+800K and a 500K+HAL. Energy scatter plots for these three models are shown in Fig.~\ref{fig:PEG_E_scatter} demonstrating that the errors on the HAL-found configurations are large for both the 500K and 500K+800K fits, despite the fact that the these HAL-found configurations are also low in energy! Only by including the HAL configurations in the training database can the errors on these configurations be reduced as shown in Table.~\ref{tbl:PEG_errors}. Inspection of the HAL generated structures exposes a shared characteristic: most of them contain (double) hydrogen bonding across the polymer an example of which is shown in Fig.~\ref{fig:PEG_E_scatter}. Such hydrogen-bond formation is a rare event in this system, because only the two ends of the molecule are capable of hydrogen bonding. It is difficult to find these configurations using regular MD (even when using elevated temperatures), whereas HAL finds them easily.

\begin{table*}
\centering
\begin{tabular}{l|c|cc|cc|cc}
\hline
\hline
 & No. & \multicolumn{2}{c|}{500K} & \multicolumn{2}{c|}{500K+800K}  & \multicolumn{2}{c}{500K+HAL} \\
 & configs & E & F  &  E & F & E  & F  \\
\hline
500K train & 750 & 30.2 & 58.3     & 32.9 &  60.8      & 32.4 & 59.6  \\
500K test & 250 & 49.2 & 79.3     & 48.8 & 76.7        & 41.6 & 71.0 \\
\hline
800K train & 250 & - & -          & 40.0 &  76.4       & - &  - \\ 
800K test & 250 & 72.7 & 187.2     & 67.6 & 107.7       & 67.9 &  102.6 \\
\hline
HAL & 200 & 310.9$^*$ &  427.2$^*$        & 311.9$^*$ &  404.6$^*$     & 47.8$^\dagger$ & 63.4$^\dagger$  \\
\hline
\hline
\end{tabular}\\
\caption{Train and test errors for energies (E) in meV and forces (F) in meV/\si{\angstrom} for the 500K, 500K+800K and 500K+HAL databases using ACE. $^\dagger$ is train error. $^*$ is test error.} 
\label{tbl:PEG_errors}
\end{table*}

\subsubsection{PEG($n$=200): bulk density}

\begin{figure*}
    \centering
    \includegraphics[width=1.0\textwidth]{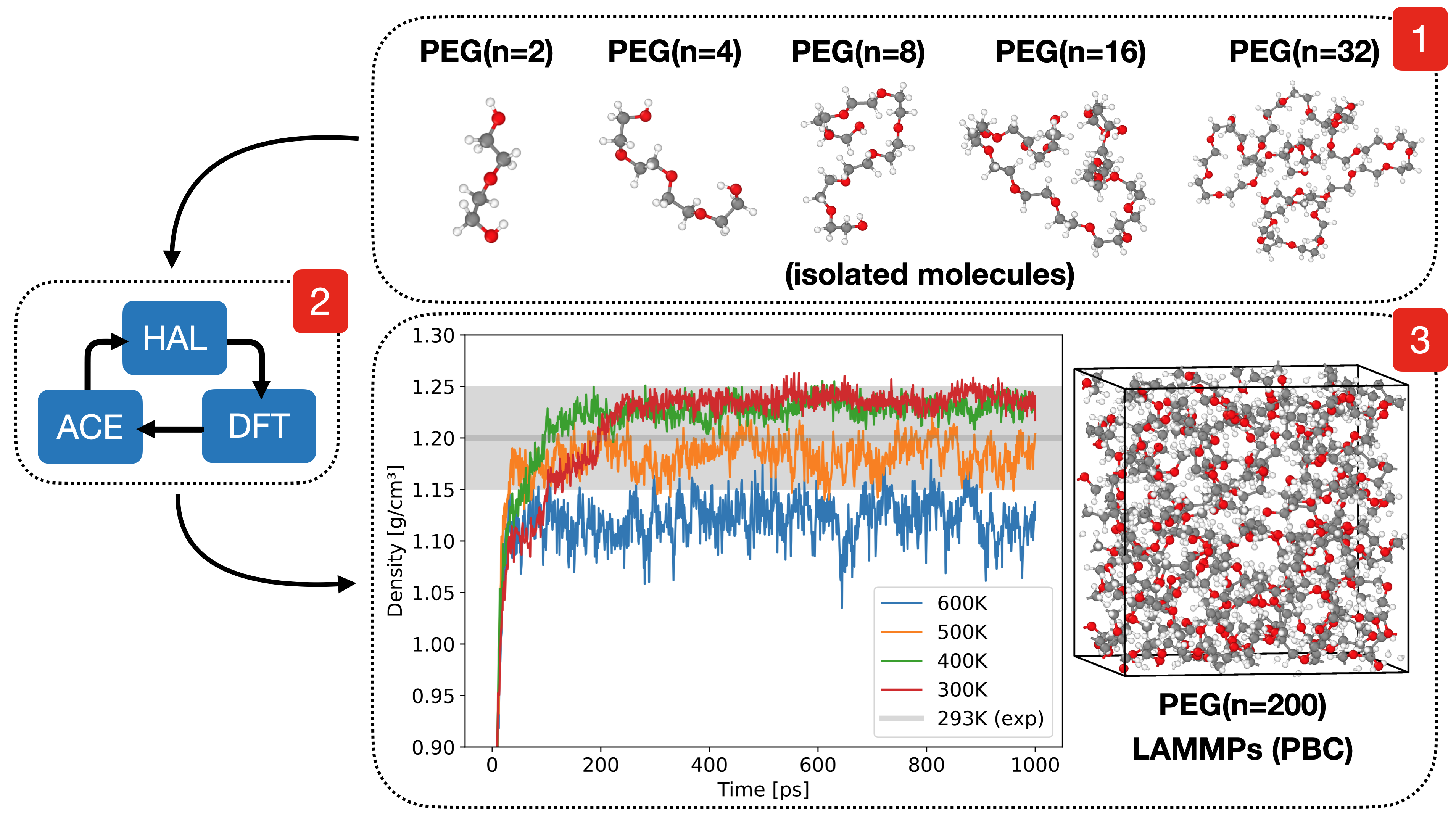}
    \caption{HAL protocol for building linear ACE PEG model accurately determining PEG($n$=200) density within experimental accuracy of 1.2 g/cm$^{3}$ at 297K (shaded area) \cite{PEG_density_reference}. Training database only included small polymers ranging from $n$=2 to $n$=32 in isolation.}
    \label{fig:PEG_comb}
\end{figure*}

As a final investigation the density of a PEG($n$=200) polymer containing 1400 atoms is determined using an ACE model fitted to a HAL generated PEG training database containing polymer sizes ranging from $n$=2 to $n$=32 monomer units. This database contained configurations from the previous PEG sections and extended using configurations sized $n=8$, $n=16$ and $n=32$. The training database included {\it standard} ANI MD sampled configurations at 500K including 1000 PEG(n=$4$) configurations (from the previous section), as well as 50 PEG($n$=2), 100 PEG($n$=8), 100 PEG($n$=16) and 18 PEG($n$=32) configurations. Starting from this data HAL was used to generate an extra 64 PEG($n$=16) and 91 PEG($n$=32) HAL configurations until dynamics was deemed stable. The linear ACE basis used for the regression task was identical to the ACE in the previous section on PEG($n$=4), and any force components with greater than 20 eV/\si{\angstrom} were excluded from the fit. 

Using the ACE model a PEG($n$=200) polymer was simulated in LAMMPS \cite{LAMMPS} with the PACE evaluator pair style with periodic boundary conditions. Since the training database only contained small polymers segments in vacuum this periodic simulation demonstrates a large degree of extrapolation to configurations far away from the training database. Furthermore, the DFT code used to evaluate the training energies and forces does not support periodic boundary conditions making DFT simulation of the 1400 atom PEG($n$=200) simulation box not just computationally infeasible, but practically impossible in this case.

The resulting linear ACE model was timed at 220 $\mu$s/atom/core per MD step. LAMMPs NPT simulations were performed at 1 bar using a 1 fs timestep at 300K, 400K, 500K and 600K. The recorded density as a function of simulation time is plotted in Fig.~\ref{fig:PEG_comb}. Using the last 500 ps from the 300K simulation the density was determined to be 1.238 g/cm$^{3}$. This value is around 3$\%$ higher than the experimental value of 1.2 g/cm$^{3}$ \cite{PEG_density_reference}.

\section{Methods}
\subsection{Hyperactive Learning (HAL)}

The HAL potential energy $\Ehal$ as defined in Eq.~\eqref{eq:ehal} biases MD simulations during the exploration step in AL towards uncertainty by shifting the potential energy surface and assigning lower energies to configurations with high uncertainty. 
We have shown that when $\sigma$ is the standard deviation of the posterior-predictive uncertainty of energy, it can be computationally cheaply approximated by a Monte-Carlo estimate $\tilde{\sigma}$; see \eqref{eq:sigma:mc}. Likewise, the derivative of $\tilde{\sigma}$ can be computed as
\begin{equation}\label{eq:std_var_rel}
 \nabla \tilde{\sigma}= \frac{\nabla \tilde{\sigma}^2}{2 \tilde{\sigma}}   
\end{equation}
where 
\begin{equation}
\begin{split}
    \nabla \tilde{\sigma}^{2} &= \frac{2}{K}  \sum_{k=1}^{K} \left( E^{k} - \bar{E} \right) \left( \nabla E^{k} - \nabla \bar{E} \right) \\
     &= \frac{2}{K} \sum_{k=1}^{K} \left( E^{k} - \bar{E} \right) \left( \bar{F} - F^{k} \right)
\end{split}
\end{equation}
and $F^k = - \nabla E^k$, $\bar{F} = - \nabla \bar{E}$. These predictions are obtained by ensemble parameterisations  $\{\mathbf{c}_{k}\}_{k=1}^{K}$, while $\bar{\mathbf{c}}$ is the analytic mean of the posterior distribution as specified in \eqref{eq:posterior:params}. The sum over $K$ is over the ensemble or committee of models, which in this work was chosen to be linear (ACE) models. Other architectures such as neural networks ensembles may be considered in future work. This quantity in essence is a computationally cheap method of determining the gradient towards (total) energy uncertainty and may be interpreted as a conservative {\em biasing force},
\begin{equation}
    F^{\tilde{\sigma}} := \nabla \tilde{\sigma}.
\end{equation}
HAL dynamics adds this biasing force to MD in order to accelerate the generation of configurations with high uncertainty, which sets HAL apart from AL. Setting $\tau$=0 recovers standard MD dynamics, and in this sense, HAL generalizes AL. Interestingly, previous work employed a biasing force using a neural network interatomic potential \cite{schran2020committee} but biased {\em away} from uncertainty in order to stabilise the MD dynamics. 

The biasing strength $\tau$ can either be set as a constant or adapted during the HAL simulation. Controlling the biasing strength is important as too strong biasing can quickly lead to unphysical configurations, whereas low biasing generates valuable configurations at a slow rate. The adaptive biasing works by first setting $\tau_{\textrm{r}}$ and performing a burn-in period to record the magnitudes (or, norms) of $F^{\tilde{\sigma}}$ and $\bar{F}$. Typically, the burn-in period is set to the latest 100 timesteps $\delta t$. The biasing strength $\tau$ is then determined by satisfying
\begin{equation}
    \tau_{\textrm{r}}= \frac{\tau  \sum_{m=1}^{100} \| \bar{F}(t - m \delta t) \| }{ \sum_{m=1}^{100} \| F^{\tilde{\sigma}}(t - m \delta t) \| }.
\end{equation}

The new parameter $\tau_{\textrm{r}}$ is generally set between 0.05 and 0.25. It can be understood as the approximate relative average strength of the biasing force in comparison to the average force of the fitted model. Using this adaptive biasing term aids usability and tunes the biasing strength to ensure that HAL gently drives MD towards high uncertainty. The value may loosely be interpreted as the relative magnitude of the biasing force compared to the true gradient of the potential energy surface. Larger $\tau_{\textrm{r}}$ increases the biasing strength and rate at which configurations with high uncertainty are generated. In order to sample configurations at desired pressures and temperatures a proportional control barostat was added as well as a Langevin thermostat. 

\subsection{Atomic Cluster Expansion (ACE)}
The ACE model decomposes the total energy $E$ of a configuration $R$ as a sum of parameterised atomic energies, 
\begin{equation}
    E(\mathbf{c}; R) 
    = \sum_{i \in R}E_{i}(\mathbf{c}; R).
\end{equation}
The atomic energies $E_i$ are then parameterised by a linear model, $E_i = {\bf c} \cdot {\bf B}_i$, where ${\bf B}_i$ denotes the ACE basis. The present work employs a particularly simple variant, which we review briefly: Given relative atomic positions $\r_{ji} = \r_{j}-\r_{i}$ and associated chemical elements one evaluates a one-particle basis 
\begin{equation}
    \phi_{znlm}(\r_{ji}, z_j) = 
    \delta_{z z_j} R_{n}(r_{ji}) Y_{lm}(\hat{\r}_{ji}),
\end{equation}
followed by a pooling operation resulting in features
\begin{equation} 
    A_{iznml} = \sum_{j} \phi_{znlm}(\r_{ji}, z_j),
\end{equation}
that are denoted the {\em atomic basis} in the context of the ACE model. Taking a $\nu$ order (tensor) product results in many-body correlation functions incorporating ($\nu$+1) body-order interactions,
\begin{equation}
    \mathbf{A}_{i\mathbf{znlm}} = 
    \prod^{\nu}_{t=1} A_{i z_{t} n_{t} l_{t} m_{t}}.
\end{equation}
The $\mathbf{A}$-basis is a complete basis of permutation-invariant functions but does not incorporate rotation or reflection symmetry. An isometry invariant basis $\mathbf{B}$ is constructed by averaging over rotations and reflections. Representation theory of the orthogonal group $O(3)$ shows that this can be expressed as a sparse linear operation and results in 
\begin{equation}
    \mathbf{B}_{i} = \mathbf{C} \mathbf{A}_{i},
\end{equation}
where $\mathbf{C}$ contains generalised Clebsch-Gordan coefficients; we refer to \cite{drautz2019atomic,DUSSON2022110946} for further details. 

A major benefit of the linear ACE model is that the computational cost of evaluating a site energy $E_i$ scales only linearly with the number of neighbouring atoms, as well as with the  body order $\nu+1$. 

\subsection{(Bayesian) Linear Regression}
The parameters of linear ACE models are fitted by solving a linear regression problem. The associated squared loss function $L(\mathbf{c})$ to be minimised over configurations $R$ in training set $\boldsymbol{R}$ with corresponding (DFT) observations for energy $\mathscr{E}_{R}$, forces $\mathscr{F}_{R}$ is 

\begin{equation}\label{eq:lsq}
\begin{split}
    L(\mathbf{c}) = \sum_{R \in \boldsymbol{R}} ( w_{E} | E(\mathbf{c}; R) - \mathscr{E}_{R} |^{2} \\
    + w_{F} | F(\mathbf{c}; R) - \mathscr{F}_{R} |^{2})
\end{split}
\end{equation}
where $w_{E}$ and $w_{F}$ are weights specifying the relative importance of the DFT observations. When fitting materials a third term is added $w_{V} | V(\mathbf{c}; R) - \mathcal{V}_{R} |^{2}$ referring to the virial stress components of the configuration $R$. This minimisation problem can be recast in the generic form 
%
%
%


\begin{equation}\label{eq:lsq:design}
      \underset{\mathbf{c}}{\arg\min}  \, \, \| \mathbf{y} - \mathbf{\Psi} \mathbf{c}  \|^{2} + \eta  \| \mathbf{c} \|^{2},
\end{equation}
where \mse{\remove{$\mathbf{\Psi}$}$\mathbf{\Psi} \in \mathbb{R}^{N_{\rm obs}\times N_{\rm basis}}$} is the design matrix and \mse{the observation vector}\mse{\remove{$\mathbf{y}$}$\mathbf{y}\in \mathbb{R}^{N_{\rm obs}}$} collects the observations to which the parameters are fitted. \mse{Entries in the design matrix and the observations vector corresponding to force observations and observations of virials are scaled by a factor of $w_E/w_F$ and $w_V/w_F$, respectively, to account for the relative weighting of the penalty terms in \eqref{eq:lsq}.} Here, we also added a Tychonov regularisation with regularisation parameter $\eta>0$ which is commonly determined through a model selection criterion such as cross-validation.

This linear regression model can be cast in a Bayesian framework by specifying a prior distribution $p({\bf c})$ over the regression parameters, and an (additive) probabilistic \mse{\remove{error model $\boldsymbol{\epsilon}$, which gives rise to the generative model}error models $\epsilon^E_R, \epsilon^F_R$ which give rise to the generative model
\begin{equation}\label{eq:linear:reg:1}
\begin{aligned}
\mathscr{E}_{R} &= E(\mathbf{c}; R) + \epsilon^E_R, \\ 
\mathscr{F}_{R} &= F(\mathbf{c}; R) + \epsilon^F_R,  
\end{aligned}
\end{equation}
for $R \in \boldsymbol{R}$. This generative model can be written in short-hand form as 
}
\begin{equation}
    \mathbf{y} = \mathbf{\Psi} \mathbf{c} + \boldsymbol{\epsilon},
\end{equation}

noise model gives rise to the likelihood function
%
%
\begin{equation}
\begin{split}
p(\mathbf{y} | \mathbf{R}, \mathbf{c}, \lambda) = \left ( \frac{\lambda}{2\pi} \right )^{N_\textrm{obs}/2} \exp \left \{  -\frac{\lambda}{2} \| \mathbf{y} - \mathbf{\Psi} \mathbf{c} \|^{2} \right \}
\end{split}
\end{equation}
%

By restricting ourselves to a Gaussian error model, and assuming the prior to be Gaussian as well, i.e., $p({\bf c}) = \mathcal{N}({\bf c} | {\bf 0}, {\bf \Sigma}_0)$, it is ensured that the posterior distribution, $\posterior({\bf c}) = p(\mathbf{c} | {\bf R},\mathbf{y}, \lambda)$, is Gaussian with closed form expressions for both the distribution mean $\bar{\bf c}$ and variance ${\bf \Sigma}$,
\begin{equation}\label{eq:posterior:params}
\begin{split}
    \bar{\mathbf{c}} &= \lambda \boldsymbol{\Sigma} \mathbf{\Psi}^{T} \mathbf{y} \\
    \boldsymbol{\Sigma}^{-1} &= {\bf \Sigma}_0^{-1} + \lambda \mathbf{\Psi}^{T} \mathbf{\Psi}.
\end{split}
\end{equation}
In the context of this work, having closed form expressions for both these quantities is desirable as it (i) allows for conceptual easy and fast generation of independent samples $\{{\bf c}^k\}_{k=1}^K$ from the posterior distribution, and (ii) allows for a parametrisation of the fitted model with the exact mean, $\bar{\bf c}$, of the posterior distribution. 

In what follows we briefly describe two Bayesian regression techniques, Bayesian Ridge Regression (BRR), which we use to produce Bayesian fits during the HAL data generation phase, and the \mse{\remove{more expensive}computationally more costly} Automatic Relevance Determination (ARD), which we typically use to obtain a final model fit after the data generation is complete. 

\subsection{Bayesian Ridge Regression (BRR)}

In Bayesian Ridge Regression the covariance of the prior is assumed to be isotropic, i.e.,
\begin{equation}
    p( \mathbf{c} | \alpha ) = \mathcal{N} (  \mathbf{c} | {\bf 0}, \alpha^{-1}  \mathbf{I} ),
\end{equation}
for some hyper-parameter $\alpha>0$, the precision of the prior distribution.

Under this choice of prior, the logarithm of the posterior distribution takes the form
\begin{equation}
    \ln \posterior({\bf c} )=  - \frac{\lambda}{2} \| \mathbf{y} - \mathbf{\Psi} \mathbf{c}  \|^{2} - \frac{\alpha}{2}  \| \mathbf{c} \|^{2} + \textrm{C}, 
    \label{eq:loglikelihood}
\end{equation}
where $C$ is some constant. Thus, maximising the (log-)posterior for this choice of prior, is equivalent to solving the regularised least square problem Eq.~\ref{eq:loglikelihood} with ridge penalty $\eta = \alpha/\lambda$. This shows that the prior naturally gives rise to a regularised solution, keeping coefficient parameters small.

The determination of the hyper-parameters $\alpha$ and $\lambda$ in BRR is achieved by optimising the marginal log likelihood also known as evidence maximisation \cite{EM}. 
One first defines the evidence function as
\begin{equation}
    p(\mathbf{y} | \alpha, \lambda) = \int p(\mathbf{y} | \mathbf{c}, \lambda) p(\mathbf{c}|\alpha) d\mathbf{c}
\end{equation}
which marginalises out the coefficients $\mathbf{c}$ and describes the likelihood of observing the data given the hyperparameters $\alpha$ and $\lambda$. Using the previously defined definitions the evidence function can be expressed as 
\begin{equation}
\begin{split}
      p(\mathbf{y} | \alpha, \lambda) =  \left ( \frac{\lambda}{2\pi} \right )^{N_\textrm{obs}/2}  \left ( \frac{\alpha}{2\pi} \right )^{N_\textrm{basis}/2} \\
      \int \exp  \left \{ - \frac{\lambda}{2} \| \mathbf{y} - \mathbf{\Psi} \mathbf{c}  \|^{2} - \frac{\alpha}{2}  \| \mathbf{c} \|^{2} \right \}
\end{split}
\end{equation}
where $N_\textrm{basis}$ is the dimensionality of $\mathbf{c}$. Completing the square in the exponent and taking the log gives rise to the marginal log likelihood
\begin{equation}
\begin{split}
\ln p(\mathbf{y} | \alpha, \lambda) =  \frac{N_\textrm{basis}}{2} \ln \alpha + \frac{N_\textrm{obs}}{2} \ln \lambda  \\
- \frac{\lambda}{2} \| \mathbf{y} - \mathbf{\Psi} \mathbf{c}  \|^{2}  - \frac{\alpha}{2}  \| \mathbf{c} \|^{2} + \\
\frac{1}{2} \ln \| \boldsymbol{\Sigma} \| - \frac{N}{2} \ln (2\pi)
\end{split}
\end{equation}
which can be maximised with respect to $\alpha$ and $\lambda$ in order maximise the marginal likelihood and obtain the statistically most probably likely solution given the basis and data. 





\subsection{Automatic Relevance Determination (ARD)}\label{sec:methods:ard}
Automatic Relevance Determination (ARD)
modifies BRR by relaxing the isotropy of the prior and assigning a hyperparameter $\alpha_{i}$ to independently regularise each coefficient $c_{i}$. The corresponding prior is given by 
\begin{equation}
\begin{split}
p( \mathbf{c} | {\bm \alpha} ) &= \mathcal{N} ( \mathbf{c} | {\bf 0,} \mathcal{A}^{-1}  ) \\ 
\mathcal{A} &= {\rm diag}(\alpha_{1}, ..., \alpha_{N_\textrm{basis}} ).
\end{split}
\end{equation}
This prior determines the relevance of each parameter $c_{i}$, or basis function, which effectively results in a feature selection. Basis functions are ranked based on their relevance and are pruned if determined irrelevant, in turn producing a sparse solution. 
In practice, sparse models obtained through ARD often yield better generalisation than BRR. Using ARD requires the specification of a threshold parameter $\alpha^{\prime}$ setting the minimum relevance of basis functions included in the fit. Adjusting this parameter controls the balance between accuracy and sparsity of the model.

\subsection{Posterior Predictive Distribution}

A key property of the Bayesian approach is that it provides a way to quantify uncertainty \mse{of model predictions} in terms of the posterior-predictive distribution, which accounts both for parameter uncertainty as given by the posterior distribution as well as uncertainty due to observation error.

\section{Data Availability}

The data will be made available at time of publication. 

\section{Code Availability}

The HAL code will be made available at time of publication. 

\section{Acknowledgements}

GC and CvdO acknowledge the support of UKCP grant number EP/K014560/1. CvdO would like to acknowledge the support of EPSRC (Project Reference: 1971218) and Dassault Systèmes UK. CO acknowledges support of the NSERC Discovery Grant (IDGR019381) and the NFRF Exploration Grant GR022937.

\section{Author Contributions}

CvdO developed efficient uncertainties for ACE models. MS conceived the idea of uncertainty-biasing of AL exploration. CvdO led the implementation of the HAL framework. DK helped generate data. CvdO wrote the first version of the manuscript. All authors discussed the theory and results and edited the manuscript.

\section{Competing Interests}

The authors declare no competing interests. 

\bibliographystyle{unsrt}
\bibliography{refs}

\end{document}